\documentclass[peerreview]{IEEEtran}
\IEEEoverridecommandlockouts
\usepackage{cite}
\usepackage{amsmath,amssymb,amsfonts}
\usepackage{algorithmic}
\usepackage{graphicx}
\usepackage{textcomp}
\usepackage{xcolor}
\def\BibTeX{{\rm B\kern-.05em{\sc i\kern-.025em b}\kern-.08em
    T\kern-.1667em\lower.7ex\hbox{E}\kern-.125emX}}
\begin{document}

\title{Influence of Neighborhood on the Preference of an Item in eCommerce Search\\
}


\author{\IEEEauthorblockN{1\textsuperscript{st} Saratchandra Indrakanti} \\
\IEEEauthorblockA{\textit{eBay Inc.} \\
San Jose, California, USA \\
sindrakanti@ebay.com} \\ 
\and
\IEEEauthorblockN{2\textsuperscript{nd} Svetlana Strunjas} \\
\IEEEauthorblockA{\textit{eBay Inc.} \\
San Jose, California, USA \\
sstrunjas@ebay.com} \\ 
\and
\IEEEauthorblockN{3\textsuperscript{rd} Shubhangi Tandon}  \\
\IEEEauthorblockA{\textit{eBay Inc.} \\
San Jose, California, USA \\
shtandon@ebay.com} \\ 
\and 
\IEEEauthorblockN{4\textsuperscript{th} Manojkumar Kannadasan}  \\
\IEEEauthorblockA{\textit{eBay Inc.} \\
San Jose, California, USA \\
mkannadasan@ebay.com}
}


\maketitle
\IEEEpeerreviewmaketitle

\begin{abstract}
Surfacing a ranked list of items for a search query to help buyers discover inventory and make purchase decisions is a critical problem in eCommerce search. Typically, items are independently predicted with a probability of sale with respect to a given search query. But in a dynamic marketplace like eBay, even for a single product, there are various different factors distinguishing one item from another which can influence the purchase decision for the user. Users have to make a purchase decision by considering all of these options. Majority of the existing learning to rank algorithms model the relative relevance between labeled items only at the loss functions like pairwise or list-wise losses \cite{burges2010ranknet,cao2007learning,xia2008listwise}. But they are limited to point-wise scoring functions where items are ranked independently based on the features of the item itself. In this paper, we study the influence of an item's neighborhood to its purchase decision. Here, we consider the neighborhood as the items ranked above and below the current item in search results. By adding \textit{delta features} comparing items within a neighborhood and learning a ranking model, we are able to experimentally show that the new ranker with \textit{delta features} outperforms our baseline ranker in terms of Mean Reciprocal Rank (MRR) \cite{Craswell2009}. The ranking models with proposed \textit{delta features} result in $3-5\%$ improvement in MRR over the baseline model. We also study impact of different sizes for neighborhood. Experimental results show that neighborhood size $3$ perform the best based on MRR with an improvement of $4-5\%$ over the baseline model.
\end{abstract}


\section{Introduction}\label{sec-intro}
Search ranking is a widely studied problem in both academia and industry. A lot of research has been performed in improving the learning to rank frameworks employed in different applications like web search, eCommerce search, question answering systems, recommendation systems \cite{liu2009learning,li2011short}. In eCommerce, given a query $q$, a typical search system retrieves all items $I_n \in I$ matching the query, ranks the items based on a ranking function $f(q, I_n)$ and returns the top $N$ documents. The ranking function $f(q, I_n)$ usually provides the probability of click or sale \cite{radlinski2005query,joachims2005accurately} of an item, independent of other items in $I$, which in turn is used to sort items. 

On the other hand, shoppers on eCommerce sites tend to compare and evaluate the list of items presented in search results, considering different options/selections available while making their purchase decision. This is somewhat different from web search, where the goal is to satisfy a single informational need. The comparative evaluation of eCommerce search results indicates that a shopper's perception of an item may be influenced by neighboring items presented along with it in the ranked results. However, the ranking functions learnt and applied in most eCommerce sites today score items independently and do not take the neighborhood into consideration. To that end, in this paper we study the influence of neighboring items on a user's preference of a given item in eCommerce search. Specifically, we aim to evaluate if incorporating the knowledge of neighborhood can help us better predict the preference of an item in the context of eCommerce search. 

For learning the ranking function, training data can be collected in 2 ways. One approach is to obtain human judged labels for items matching a query, to annotate a binary decision of relevant or not for a given item \cite{radlinski2005query}. Second approach is to extract implicit relevance feedback based on user behavior logs \cite{agichtein2006learning,cohen1998learning,Joachims:2002:OSE:775047.775067}. In web search as well as in eCommerce search, one of the widely used relevance feedback is clicks. In addtion to that, eCommerce search systems have the advantage of using more relevance feedback signals like bids, add to carts, purchases, revenue etc \cite{KarmakerSantu:2017:ALR:3077136.3080838}. The basic assumption in implicit relevance feedback is, users scan the items in \textit{top-down} manner. Existing literature study the impact of items that were viewed and not clicked as negative samples in relevance feedback \cite{Joachims:2002:OSE:775047.775067}. Other studies have focused on the impact of a document's relevance based on the documents ranked above it with the focus on search result diversity \cite{zhu2014learning,Agrawal:2009:DSR:1498759.1498766}.  
In this paper, we study the effect of the neighboring items, i.e. items ranked above and below a particular item $I_n$ on the preference of $I_n$ in eCommerce search results. To evaluate the impact, we quantify neighborhood by means of features that compare items ranked at different positions above and below the current item. These comparative features are denoted as \textit{delta features}. 

Our study highlights different \textit{delta features} we tried on top of our current baseline model, and the improvements in offline metrics they result in. We also evaluate the effect of different neighborhood sizes $m$ used in constructing the \textit{delta features}, and experimentally show that the neighborhood of an item has an impact on the item's preference in the ranked results through offline metrics.

The rest of the paper is organized as follows. Section \ref{sec-relatedwork} discusses some of the related work in the literature. In Section \ref{section-approach} we describe our methodology. In Section \ref{sec-experiments} we describe our datasets and experiments. We summarize our work and discuss possible future research in Section \ref{sec-summary}. 
\section{Related Work}\label{sec-relatedwork}
Lichtenstein et. al presented some early work on how people make decisions under uncertainty in \cite{lichtenstein1971reversals}, where the key insight is that the decisions are different when choices are presented separately vs. when they are presented together. Importance of a context (neighborhood) for a given item to its clickability has been extensively researched in the past. Previous studies of users' clicks as implicit feedback in search found out that clicking decision on a web document is affected by both rank and other documents in the presentation \cite{ joachims2008clickthrough}, \cite{ joachims2007accuracy}. Craswell \textit{et al.} \cite{ craswell2008cascade} introduced \textit{the cascade click model} where the probability of click for a given document at a given rank is influenced by probability of click for documents at higher ranks.

Dupret \textit{et al.} \cite{ dupret2008browsing} introduced a new browsing behavior model, where the probability of a subsequent click for a given document is affected by a distance between that document and the most recently clicked document. The probability gets lower if the previously clicked document is further away, i.e. if a user has to scroll through numerous irrelevant documents. Our approach extends this research to model preference of items in e-commerce search.
\section{Our Approach}\label{section-approach}
Our hypothesis is that whenever users make a decision to buy an item on an eCommerce platform, it is not in isolation. The decision is made by comparing the item to other items in its vicinity. Most ranking models use a single item's features to determine the probability of sale. To understand how the neighboring items affect an item's preference by a user, we define \textit{delta features} that represent how the item differs from it neighboring items. 

We focus on features that could be potentially distinguishing factors of an item and those that can identify user behavior. Since we want to model user behavior, these features are derived from elements users are likely to see on the search results page when making a purchase, for e.g. shipping time, product title, product price etc. We identified the set of features which users are likely to perceive while buying an item as the candidate set $\left( F : f_{1}, f_{2} …. f_{n} \right) $ from which we can generate \textit{delta features}.

\begin{figure}[h]
\small
\includegraphics[scale=0.4]{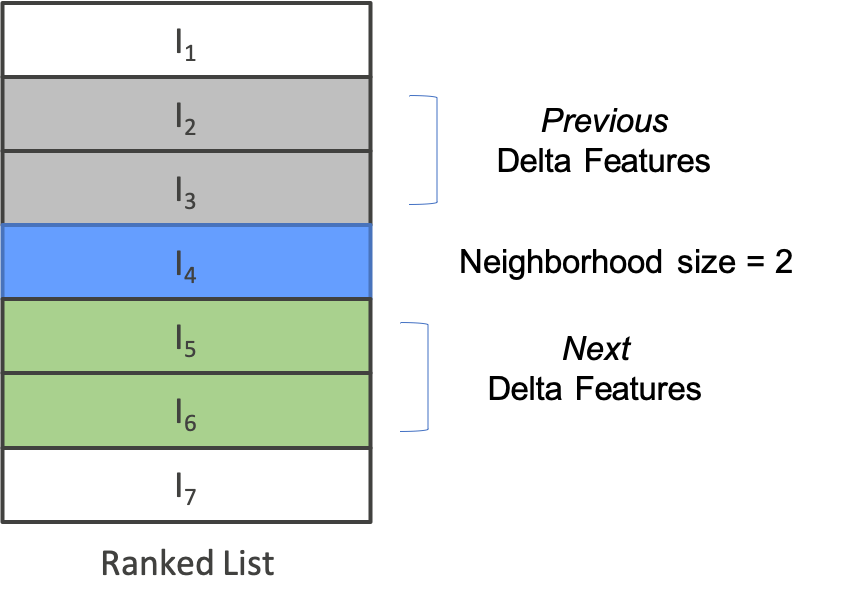}
\setlength{\belowcaptionskip}{-10pt}
\caption{Illustration of previous and next \textit{delta features} constructed based on a ranked list of items. Here the neighborhood size is 2.}
\label{fig:delta_features}
\end{figure}

We experiment with three different neighborhood sizes ( size = 1, 3,5 )  to study how the influence of the delta features changes as the neighborhood size changes.  For each of these candidate features $F$, we generated two types of \textit{delta features} each, namely \textit{next} and \textit{prev}; \textit{next} represents the \textit{delta features} based on the items ranked below the current item, while \textit{prev} represents the \textit{delta features} based on the items ranked above the current item. Fig \ref{fig:delta_features} represents an example of a neighborhood of size 2. For the item $I_4$, \textit{next} features are calculated by comparing features of $I_4$ with $I_5$ and $I_6$. Similarly, \textit{prev} features are calculated by comparing features of $I_4$ with $I_2$ and $I_3$. Note that neighborhood size refers to the number of items considered in computing the \textit{delta features} above and below the current item. The \textit{delta features} are denoted as,
\begin{align*}\label{eq:1}
 D &: \big[ d_{1m\_prev},\; d_{1m\_next}, \; d_{2m\_prev},\; d_{2m\_next},  \\
& \ldots ,\;  d_{nm\_prev},\;  d_{nm\_next} \big]
\end{align*}

where $m$ represents the neighborhood size. We further define a distance weighted decay function $\gamma(j)$, where $j$ is the number of positions a neighbor is away from the current item. $\gamma(j)$ captures varying distance adjusted contributions to the delta feature by different neighbors, based on the intuition that items that are farther may have a different influence on a users' perception of an item than a closer one.
There are three different categories of delta features defined : 
\begin{enumerate}
\item \textbf{\textit{Numerical Delta Features}} :  Numerical delta features are defined as the difference between the previous/next item's features and the current item's features:

	\[ D_{km\_prev} = \frac{1}{m} *\sum_{j=1}^{m} \frac{f_{k-j} - f_{k}}{\gamma(j)} \]
	\[ D_{km\_next} = \frac{1}{m} *\sum_{j=1}^{m} \frac{f_{k+j} - f_{k}}{\gamma(j)} \]

\item \textbf{\textit{Categorical Delta Features}} :  For categorical features with discrete values, the delta features are defined as the distance weighted average of matching discrete feature values occurring in the neighborhood of the current item. This can be represented as:

	\[ D_{km\_prev} = \frac{1}{m} *\sum_{j=1}^{m} \frac{diff(f_{k-j}, f_{k})}{\gamma(j)} \]
	\[ D_{km\_next} = \frac{1}{m} *\sum_{j=1}^{m} \frac{diff(f_{k+j}, f_{k})}{\gamma(j)} \]

where $diff(a,b) = 1$ if $a=b$, and $0$ otherwise.  
Note that, boolean delta features are a special case of categorical ones, where there are only 2 possible feature values.
 
\item \textbf{\textit{Vector based Delta Features}}  : Delta features can be computed based on vector based representations of items. For instance, item embeddings learnt based on specific properties and subsequent user interactions can be used as representations to effectively capture similarities and differences between items. 
 
	\[ D_{km\_prev} = \frac{1}{m} *\sum_{j=1}^{m} \frac{Vdiff(v_{k-j}, v_{k})}{\gamma(j)} \]
	\[ D_{km\_next} = \frac{1}{m} *\sum_{j=1}^{m} \frac{Vdiff(v_{k+j}, v_{k})}{\gamma(j)} \] 
	
where $v_{k}$ is the vector representing the item at position $k$  and $Vdiff(\alpha,\beta)$ is a distance measures between vectors $\alpha$ and $\beta$ of the same dimensionality. A measure such as cosine similarity may be used for this purpose where $Vdiff(\alpha,\beta)$ can be defined as $1-cos(\alpha,\beta)$.
 
%
%
\end{enumerate}

\section{Experiments}\label{sec-experiments}
We build several offline ranking models with varying neighborhood sizes and selection of delta features to evaluate the incremental improvement produced by these features in the performance of the ranking models, and subsequently observe the effect of neighborhood on the preference of an item. In this section, we will describe the dataset used, the various feature sets employed in the experiments that follow, and the models built as part of the experiments. 

\subsection{Dataset, Features and Experiment Setting}\label{subsection-datasetexpsetting}
We conduct our ranking experiments on a large-scale dataset sampled from eBay search logs. The dataset consists of about 20000 unique search queries sampled based on user search sessions which resulted in an item's sale, along with the ranked list of top items impressed for the query. The labels for the items in the dataset are obtained via implicit relevance feedback. In this paper, we consider the sale of an item as the target. We constructed \textit{delta features} as described in Section \ref{section-approach} based on features that are perceivable by the users such as price, popularity and retail standards associated with the item. While, embedding based delta features can be constructed using item embeddings, we limit delta features to either numerical or categorical in the experiments that follow. Further, we use a distance weighted decay function $\gamma(j) = 1$ in constructing delta features. In other words, we treat farther neighbors the same as closer ones while computing delta features. $80\%$ of the dataset was used for training and $20\%$ for validation. 

We trained several learning to rank models on the dataset described above. We use the state-of-the-art LambdaMART model \cite{burges2010ranknet} for our experiments. The baseline model, \textbf{\textit{Model\_Base}} is trained on the same dataset without any \textit{delta features}. \textit{Model\_Base} is the production ranking model for eBay. The proposed ranking models use features from \textit{Model\_Base} and \textit{delta features}. We train ranking models with different neighborhood sizes and different neighborhood types namely, \textit{prev} and \textit{next}. We experimented with 3 neighborhood sizes in this paper, $m=1, 3, 5$. We trained three different models for each neighborhood size, $m$:
\begin{enumerate}
	\item \textbf{\textit{Model\_Prev\_Wm}} : Models with \textit{prev delta features}, calculated based on items ranked above the current item
	\item \textbf{\textit{Model\_Next\_Wm}}  : Models with \textit{next delta features}, calculated based on items ranked below the current item
	\item \textbf{\textit{Model\_Prev\_Next\_Wm}} : Models with \textit{prev} and \textit{next delta features}, calculated based on items ranked above and below the current item
\end{enumerate}
The hyperparameters are tuned based on \textit{Model\_Base} and the same parameters are used to train all the proposed ranking models with \textit{delta features}.  

\subsection{Results}\label{subsection-results}
We trained models with both previous and next delta features constructed based on neighborhood sizes 1, 3 and 5 respectively. The trained models were evaluated offline on the test dataset with the aim being observing incremental ranking improvements to the models introduced by delta features. Mean reciprocal sale rank (MRR) was chosen as the metric to evaluate and compare the performance of the various models relative to the baseline model \textit{Model\_Base}. MRR, in this case captures the first result that involves an item sale. We employed MRR as the evaluation metric to capture the notion of preference in a ranked list via sale of an item. 



\begin{figure}[h]
\small
\includegraphics[width=0.5\textwidth]{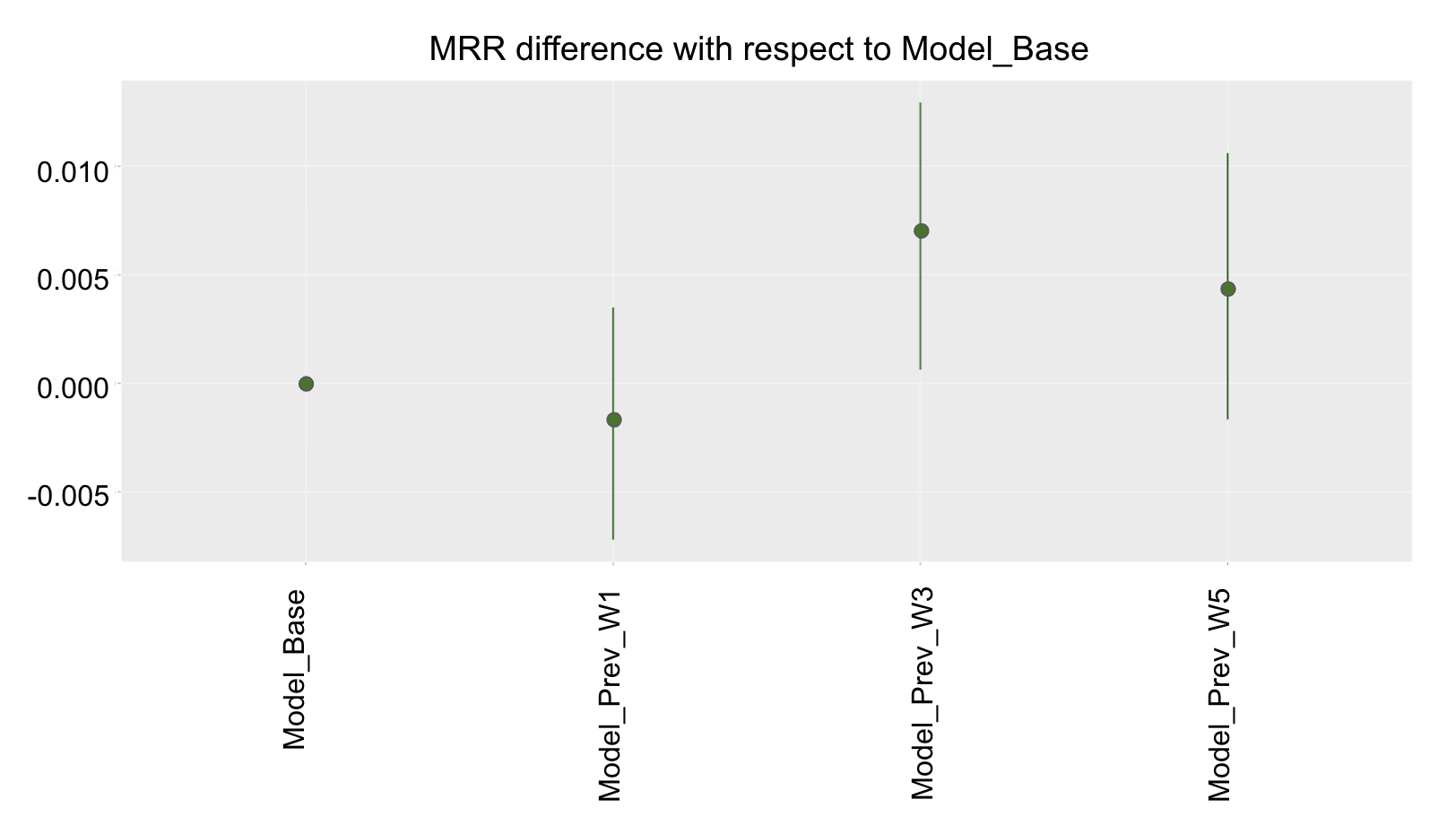}
\setlength{\belowcaptionskip}{-10pt}
\caption{MRR difference with respect to \textit{Model\_Base} for neighborhood sizes 1, 3 and 5 using \textit{prev} features.}
\label{fig:prev_w135}
\end{figure}

\begin{figure}[h]
\small
\includegraphics[width=0.5\textwidth]{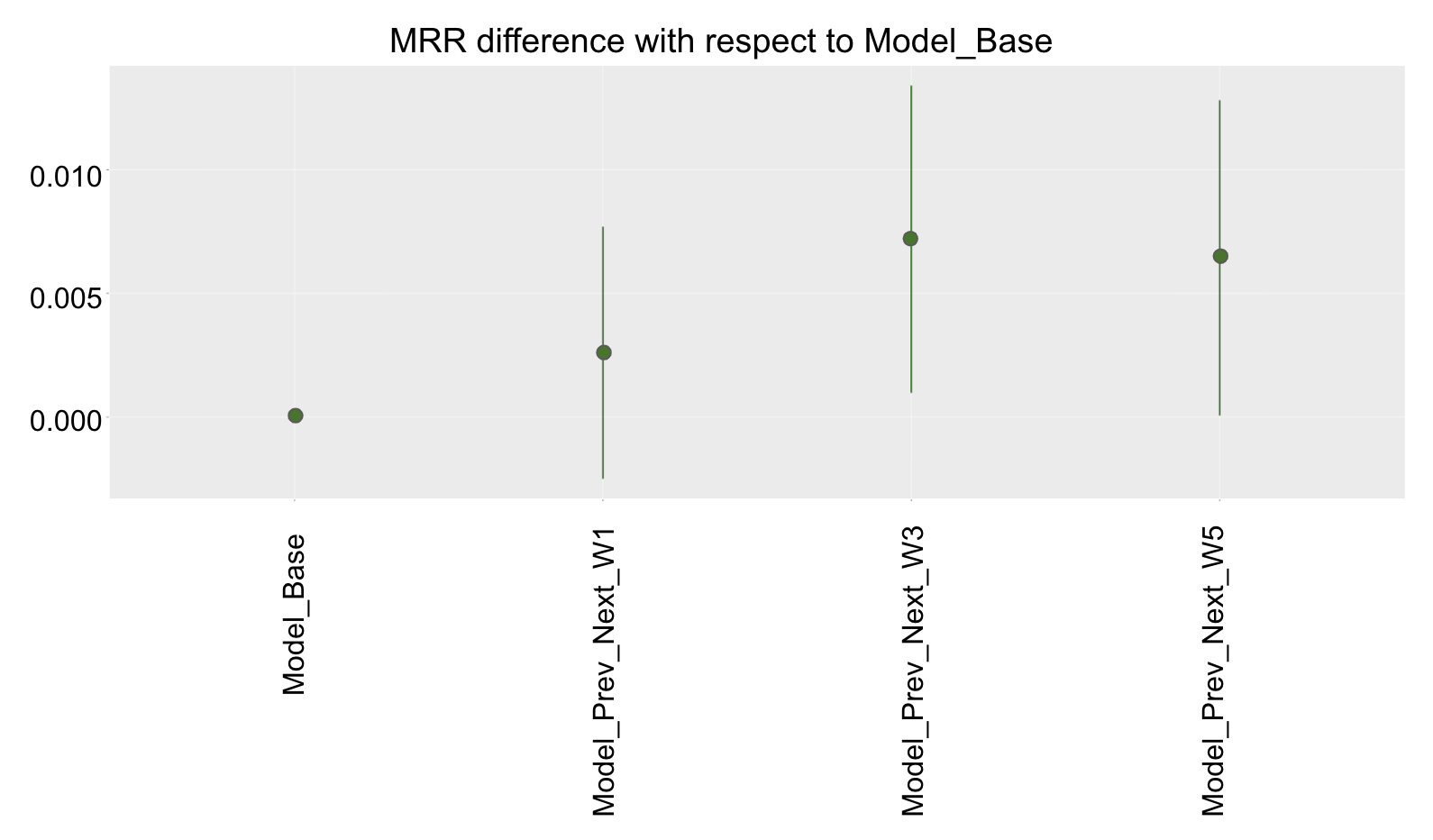}
\setlength{\belowcaptionskip}{-10pt}
\caption{MRR difference with respect to \textit{Model\_Base} for neighborhood sizes 1, 3 and 5 using both \textit{prev\_next} features. }
\vspace{1mm}
\label{fig:prev_next_w135}
\end{figure}

The \textit{prev} and \textit{next} features which capture the neighborhood above and below an item in the ranked list of results, show significant improvements in MRR compared to the baseline model. The figures show MRR difference with respect to \textit{Model\_Base} and the error bars are computed using 1000 bootstrap samples of the test dataset.
 
\begin{figure}[h]
\small
\includegraphics[width=0.5\textwidth]{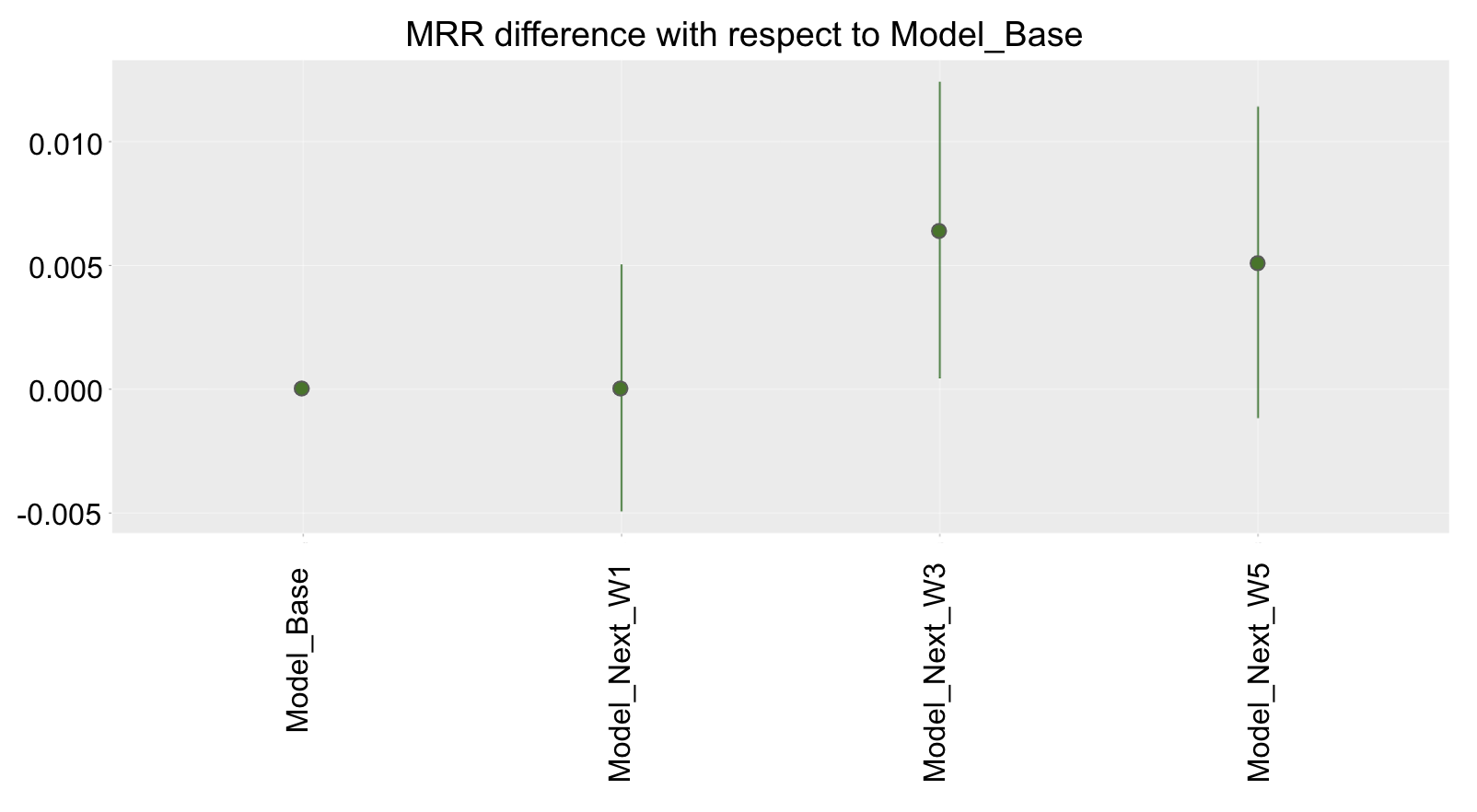}
\caption{MRR difference with respect to \textit{Model\_Base} for neighborhood sizes 1, 3 and 5 using \textit{next} features.}
\vspace{-5mm}
\label{fig:next_w135}
\end{figure}

First, we used only \textit{prev} features constructed based on neighborhood sizes 1, 3 and 5 in addition to \textit{baseline} features.  \textit{prev} features lead to MRR improvements as can be seen from Fig \ref{fig:prev_w135}, with neighborhood size 3 outperforming others. Similarly, Fig \ref{fig:next_w135} shows the relative MRR improvements when only \textit{next} features constructed based on neighborhood sizes 1, 3 and 5 in addition to \textit{baseline} features. Neighborhood size 3 leads to the most significant improvements in MRR. Further, varying  neighborhood sizes has a measurable effect on MRR, indicating that the choice of neighborhood size is an important decision. Lastly, by combining \textit{prev} and \textit{next} features on top of the \textit{baseline} features also resulted in significant improvements in MRR with neighborhood size 3, performing the best as shown in Fig \ref{fig:prev_next_w135}.

The percentage gains in MRR resulting from each of the models relative to \textit{Model\_Base} is tabulated in Table \ref{tab:percent_table}. As evident from the table, using \textit{prev\_next} features constructed using a neighborhood size, 3, results in $5.01\%$ improvement in MRR, thereby supporting the intuition that the neighborhood consisting of both items ranked above and below an item together influence preference of an item. 

\begin{table}[htbp]
\caption{Percentage change in MRR}
\begin{center}
\begin{tabular}{|l|l|l|l|}
\hline
\textbf{Neighborhood size} &\textbf{prev}  & \textbf{next} & \textbf{prev\_next} \\ \hline
1                 & -1.32 & 0.07 & 1.81       \\ \hline
\textbf{3}                 & 4.65  & 4.45 & \textbf{5.01}       \\ \hline
5                 & 3.05  & 3.55 & 4.52       \\ \hline
\multicolumn{4}{l}{Percentage change in MRR relative to \textit{Model\_Base} }\\
\multicolumn{4}{l} {resulting from the various models. }
\end{tabular}
\label{tab:percent_table}
\end{center}
\end{table}

\begin{figure}[h]
\small
\includegraphics[width=0.5\textwidth]{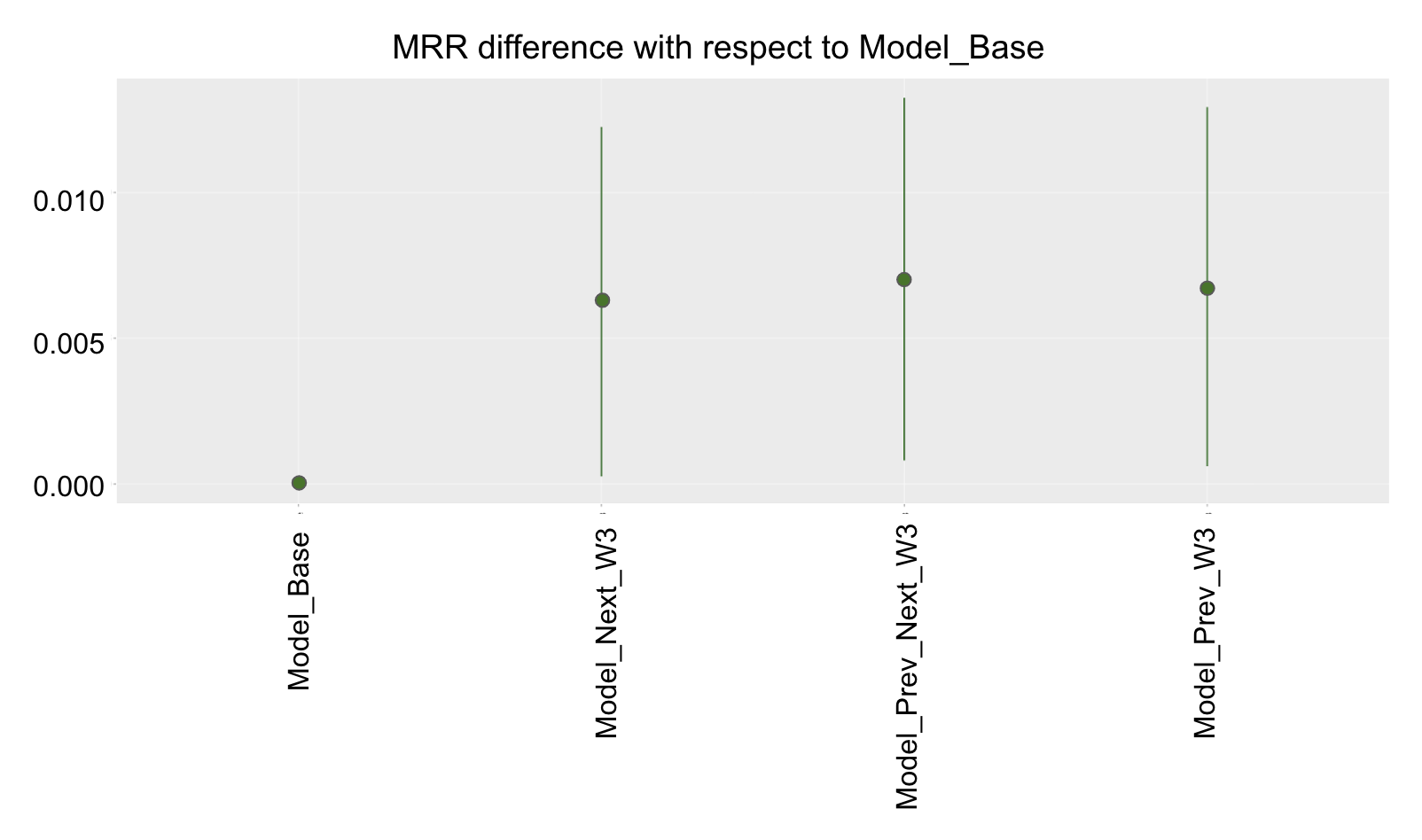}
\caption{MRR difference with respect to \textit{Model\_Base} for neighborhood size 3 using \textit{prev}, \textit{next}, and \textit{prev\_next} features.}
\vspace{-5mm}
\label{fig:prev_next_w3}
\end{figure}

Since neighborhood size 3 resulted in the most observable MRR improvements, we compared \textit{prev}, \textit{next}, and \textit{prev\_next} models trained on delta features constructed with neighborhood size 3 in addition to the \textit{baseline} features. From Fig \ref{fig:prev_next_w3} we can observe that while both \textit{prev} and \textit{next} models lead to improvements, \textit{prev\_next} models have the most pronounced MRR gains, indicating that the neighborhood of an item does influence its preference in a measurable way.  Further, the observation that larger neighborhood sizes don't necessarily contribute to more effective models suggests applying a distance weighted decay in constructing delta features. We plan to explore the effects of such a function in future work. 

\section{Summary and Future Work}\label{sec-summary}
Learning to rank techniques are widely used to learn a ranking function and furnish a ranked order of items across a broad range of domains and are a critical component of eCommerce domain specifically. In practice, items are usually ranked independently, without taking into account the influence of neighboring items on the placement of a given item. However, when users view a ranked list of items, their perception of a given item is influenced by its neighborhood. This effect is even more pronounced in eCommerce, where users have a selection of relevant items and tend to make comparative decisions. This raises the question of investigating the influence of neighborhood on the placement of an item in a ranking list. List-wise loss functions and group-wise scoring functions have been studied in literature, and methods to place an item in a ranked list based on its predecessors have been proposed. However, the influence of neighborhood on a user's perception of an item in a ranked list has been seldom investigated, specifically in the eCommerce domain. To that end, we investigated the influence of neighboring items on users' perception of a given item by studying the effect of neighborhood within a ranked list of items. 

We constructed \textit{delta features} that capture how a given item differs from those in its neighborhood in terms of attributes that can be perceived by the user on a search result page. We then trained learning to rank models based on a pairwise loss function and conversion ( sale ) as a target to study the effect of these delta features on understanding the preference of an item. By employing a feature set that consisted of the newly constructed delta features in addition to features that are already being used in models that are on site, we examined the incremental benefits of the delta features.  From our experiments, we find that delta features consistently rank high in terms of feature importance. Further, including delta features contributes positively to ranking metrics such as mean reciprocal sale rank. Including \textit{previous} and \textit{next} features outperforms using either \textit{previous} or \textit{next} individually. In addition to this, we discovered that the choice of the size of neighborhood influences the performance of these features. In summary, the key takeaways from this work are : 

\begin{itemize}
	\item The neighborhood of an item effects users' perception of it and its preference within a ranked list, specifically in eCommerce domain. Hence neighborhood must be accounted for while placing an item in a raked search result page. 
	\item The choice of the size of the neighborhood influences the performance of delta features, and subsequently the ability to model neighborhood. 
\end{itemize}

As a next step, we plan to investigate the applicability of item embeddings and the effect of introducing a distance weighted decay in the construction of delta features, as part of work focused on constructing more effective representations of neighborhoods. Another application of the learning of this work is incorporating the idea of neighborhood and delta features into ranking models. This would require designing efficient methods to determine the placement of a candidate item based on its potential neighbors, in contrast to an independent decision. Further, by identifying discriminating delta features, we may be able to understand diversity as perceived by eCommerce users. While diversity in a ranked list has been well studied in web search, a nuanced study of what attributes describe diversity in the context of eCommerce can be useful to the domain. Building up on the idea of delta features, we will study the features and attributes that can explain diversity in eCommerce as future work.

\section*{Acknowledgment}
We would like to thank Alex Cozzi for the insightful discussions and valuable guidance he provided during the course of this work. 

\bibliography{sellability.bib}{}
\bibliographystyle{IEEEtran}

\end{document}